\documentclass[fleqn,usenatbib]{mnras}
\usepackage{newtxtext,newtxmath}
\usepackage[T1]{fontenc}
\DeclareRobustCommand{\VAN}[3]{#2}
\let\VANthebibliography\thebibliography
\def\thebibliography{\DeclareRobustCommand{\VAN}[3]{##3}\VANthebibliography}
\usepackage{graphicx}	% Including figure files
\usepackage{amsmath}	% Advanced maths commands
\usepackage{xcolor}
\usepackage{pdflscape}
\newcommand{\beq}{\begin{equation}}
\newcommand{\eeq}{\end{equation}}

\def\lap{\lower.5ex\hbox{$\; \buildrel < \over \sim \;$}}
\def\gap{\lower.5ex\hbox{$\; \buildrel > \over \sim \;$}}

\title[Local Supercluster]{The Extended Local Supercluster}

\author[P. J. E. Peebles]{P. J. E. Peebles,$^{1}$\thanks{E-mail: pjep@Princeton.edu}\\
$^{1}$Joseph Henry Laboratories, Princeton University, Princeton, NJ 08544, USA\\
}

\pubyear{2021}

\begin{document}
\label{firstpage}
\pagerange{\pageref{firstpage}--\pageref{lastpage}}
\maketitle
\begin{abstract}
It has long been established but seldom noticed that we are in a region at least 170~Mpc across in which different types of galaxies show different degrees of alignment with the plane of the de Vaucouleurs Local Supercluster. While clusters of galaxies and radio galaxies at redshifts $z<0.02$ are concentrated at low supergalactic latitudes, the most luminous galaxies in the infrared, LIRGs, show little correlation with this plane. The  most luminous early-type galaxies are concentrated at low supergalactic latitudes, but similarly luminous spirals are not noticeably so. The cross-correlations of the positions of what might be considered galaxies selected for their stellar mass with positions of clusters and LIRGs offer a measure of the situation. The mean density at distance $\sim 0.5$~Mpc  from a LIRG is  comparable to the mean density at that distance from a cluster of galaxies, but the mean density 5~Mpc  from a LIRG is well below the mean density at that distance from a cluster and not much greater than the cosmic mean density. Discussion of issues arising is brief.
\end{abstract}

\section{Introduction}

Einasto, Corwin, Huchra et al. (1983, Fig. 1) and Tully (1986, 1987) pointed out that clusters of galaxies tend to be close to the plane of the de Vaucouleurs Local Supercluster in a region around us that is several hundred megaparsecs  across. Shaver and Pierre (1989) and Shaver (1990, 1991) took note of this phenomenon and added the evidence that radio galaxies at comparably large distances also tend to be at low supergalactic latitudes. 

Shaver (1991, Fig. 5) made the point that motivates the present paper. At redshifts $z<0.02$, where most clusters and radio galaxies are at low supergalactic latitudes SGB, close to the plane of the Local Supercluster, ordinary $L\sim L_\ast$ galaxies are not so noticeably concentrated to this plane. Shaver's effect has not been discussed much in the literature. Among the 25 citations to Shaver (1991) listed in NASA's Astrophysics Data System I find explicit mention of Shaver's effect in only three papers, all review articles I wrote. I continue to think the phenomenon merits closer attention; maybe it has something to teach us about the origin of cosmic structure. In this paper I add to the scanty literature on Shaver's effect by an exploration of the phenomenon in the all-sky catalogs now available, with a few thoughts about what it might mean.

We should pause to recall that G\'erard de Vaucouleurs (1953) drew attention to the tendency of  the galaxies nearer than about 30~Mpc to be close to the plane of what he first termed the Local Supergalaxy and later the Local Supercluster (or first the local super-cluster, in de Vaucouleurs 1958). More recent terms are the Virgo Supercluster, and a lobe of the Laniakea Supercluster (Tully, Courtois, Hoffman, and Pomar{\`e}de 2014). Coordinates aligned with the plane of this system still are termed supergalactic, and their worth is demonstrated by inclusion in the angular coordinate conversions provided by the NASA/IPAC Extragalactic Database. 

Community acceptance that de Vaucouleurs' phenomenon is likely to be more than a chance alignment was aided by Abell's (1961) review of Carpenter's (1961) numerical evidence of Vaucouleurs' effect. More recently  Baleisis, Lahav, Loan, and Wall (1998) found reasonably convincing evidence of the tendency of radio galaxies to be near the plane of the Local Supercluster; Lahav, Santiago, Webster, Strauss, Davis, Dressler, and Huchra  (2000) found measures of the distributions of galaxies selected in  optical and infrared surveys relative to the supergalactic plane; and B{\"o}hringer, Chon, \& Tr{\"u}mper (2021a) added the evidence that X-ray detected clusters of galaxies also tend to be close to the plane of the Local Supercluster out to distances comparable to that of the Einasto et al. (1983) sample. McCall (2014) pointed out that all the galaxies, large and small, within about 6~Mpc distance from us are remarkably close to what McCall termed the Local Sheet (a term used earlier by Tully, Shaya, Karachentsev, et al. 2008 to describe the coherent motion of the nearby galaxies away from the Local Void). McCall's sheet is slightly tilted from the plane of de Vaucouleurs' Local Supercluster. This Local Sheet includes the radio galaxy Centaurus A, at distance 3.6~Mpc, which is at  supergalactic latitude SGB$ =  -5^\circ$. But almost all galaxies at this distance are at low SGB. 

For the history of how radio astronomy became a part of physical cosmology I refer to Sullivan (2009) and chapters 35 and 36 in Goss, Hooker, and Ekers (2022). A natural part of this early research was to look for radio radiation from the concentration of galaxies around the Virgo Cluster and in the plane of de Vaucouleurs' Local Supercluster. The earliest reports of detection I have found are by Kraus and Ko (1953) at the Ohio State University and Hanbury-Brown and Hazard (1953) at the Jodrell Bank Experimental Station. Hanbury-Brown (1962) reported the detection of luminous large angular size radio sources afforded by the better angular resolution at the Mullard Radio Astronomy Observatory, and showed that they are largely at low supergalactic latitude SGB. Hanbury-Brown concluded that ``Although this evidence [of the concentration to low SGB] is far from conclusive, it is at least consistent with the model [of the Supergalaxy] we have assumed." This was an oblate concentration centered in the direction of Virgo. Hanbury-Brown's distribution of radio sources in SGB in his Fig.~7 resembles the distribution in Panel~radio$_{12}$ in Figure~\ref{hist-radio} below. The observable effect in surveys of radio sources remained subtle, however. Thus Pauliny-Toth, Witzel, Preuss, K{\"u}hr, Kellermann, Fomalont, and Davis (1978) presented evidence that extragalactic radio sources are more abundant at lower SGB and toward the Virgo cluster, and that ``The counts indicate a possible anisotropy related to the supergalactic system.'' Shaver and Pierre (1989) and Shaver (1990; 1991) had better data to make the relation of radio sources to the Local Supercluster more definite. It allowed Shaver (1991) to see that at distances out to redshift $z=0.02$, or about 85~Mpc on the present distance scale, radio galaxies tend to be at low supergalactic latitudes while ordinary $L\sim L_\ast$ galaxies are more nearly isotropically distributed. The results presented in Section~\ref{sec:angularpositions} below illustrate the present state of the evidence of this effect: different kinds of objects at the same distance from us have different degrees of concentration to low SGB.

Section~\ref{sec:data} presents conventions adopted in aid of comparisons of the distributions of supergalactic latitudes of a variety of samples of objects that have been mapped in reasonable approximations to all-sky surveys.  Understanding these results may be aided by the cross-correlations of positions of some of the samples that are presented in Section~\ref{sec:correlations}. I offer in Section~\ref{sec:conclusions}  summary comments on the results and steps that could be taken that might improve our understanding of Shaver's effect, along with brief thoughts about the possible physical significance of what is now known.

\section{Conventions}\label{sec:data}

The distribution of supergalactic latitudes of clusters of galaxies is to be compared to the distributions of galaxies selected at several different wavelength bands. The measure used in Section~\ref{sec:angularpositions} is the distribution of $\sin {\rm SGB}$, where SGB is the supergalactic latitude of an object. Baleisis, Lahav, Loan, and Wall (1998) used the same measure for the same purpose. Since equal intervals of $\sin{\rm SGB}$ mark equal elements of solid angle, the expected counts of an isotropic random process  in equal intervals of $\sin {\rm SGB}$ are independent of SGB. 

The obscuration  of objects at low galactic latitudes requires a correction to the expected distribution of $\sin{\rm SGB}$. In this paper all objects at galactic latitudes $|b|<10^\circ$ are rejected, and the effect on the distribution of $\sin {\rm SGB}$ is computed from the distribution of points placed uniformly at random in the allowed bounds of galactic latitude. If incompleteness is significant at somewhat higher galactic latitudes it tends to suppresses the counts of galaxies at high SGB. The observed distributions in sin~SGB suggest this effect is not a serious problem. 

Apart from the nearest sample to be considered, in which  galaxy distances are measured, the samples are placed in separate intervals of redshift. It is to be borne in mind that different galaxies at the same physical distance may appear in different redshift bins, in a greater degree among galaxies with larger peculiar velocities, perhaps those near clusters of galaxies. Measured redshifts are translated to physical distances by using Hubble's constant 
\beq
H_{\rm o}=70~\hbox{km}~\hbox{s}^{-1}~\hbox{Mpc}^{-1}.\label{eq:Hnot}
\eeq
The individual measurements of distances and redshifts of relatively nearby galaxies are reasonably consistent with this value of $H_{\rm o}$ (Tully, Courtois, and Sorce 2016). 

 \begin{table}
 \centering
 \caption{Sample Sizes} \label{tab:samplenumbers}
\begin{tabular}{lccc}
\hline
Redshift range &0.01 - 0.02 & 0.02 - 0.04  & 0.04 - 0.08 \\
\hline
Brightest Cluster Galaxies & 12 & 58 &  359\\
Cluster X-ray Sources & 30 & 114 & 302 \\
Radio Galaxies & 32 & 141 & 253\\
IRAS PSCz galaxies  & 61 & 523 & -- \\
2MASS  2MRS galaxies & 6515$^{\rm a}$ & 7908$^{\rm b}$ & -- \\
\hline
\multicolumn{4}{l}{$^{\rm a}$luminous enough to be detected at $z=0.02$.}\\
\multicolumn{4}{l}{$^{\rm b}$luminous enough to be detected at $z=0.04$.}\\
\end{tabular}
\end{table}

The issue to be considered is the differences of the degree of alignment to the plane of the Local Supercluster by clusters of galaxies and by galaxies selected by luminosities at different wavelengths. Catalogs of galaxies selected in different wavelength bands, if counted to large enough number densities, must contain many galaxies in common. Since it would be difficult if not impossible to sort the objects into separate interesting categories, I instead choose for a reference base the large numbers of 2MRS galaxies in the redshift catalog based on the 2MASS galaxy survey selected by apparent magnitudes $K_s$ at $2.2\mu$ (Skrutskie, Cutri, Stiening, et al. 2006; Huchra, Macri, Masters, et al. 2012; the data used here are drawn from the VizieR On-line Data Catalog).  I consider the rare most luminous of the objects in other catalogs, with spatial number densities that are comparable to a typical density of clusters of galaxies. The hope is that this may lead us to other comparably rare and perhaps comparably interesting objects. 

The translation to the mean volume number density $n$ from the number $N$ of objects in the redshift bin $0.01<z<0.02$, with the bound $|b|>10^\circ$ on the galactic latitude, and the Hubble constant in equation~(\ref{eq:Hnot}), is 
\beq
n = FN, \quad F = 5.24\times 10^{-7}\hbox{ Mpc}^{-3}.
\eeq
Doubling the redshift bounds multiplies $F$ by a factor of eight. The redshift bin $0.01<z<0.02$ includes 12 Brightest Cluster Galaxies (Lauer, Postman, Strauss, Graves, and Chisari 2014), and 6515 2MRS galaxies that are luminous enough to be detected at $z=0.02$. They translate to number densities
\beq\label{eq:densities}
\begin{split}
n_{\rm BCG} &= 6.3\times 10^{-6}\hbox{ Mpc}^{-3},  \\
n_{\rm 2MRS} &= 3.4\times 10^{-3}\hbox{ Mpc}^{-3}.
\end{split}
\eeq
The 2MRS density is comparable to that of optically selected $L\sim L_\ast$ galaxies. The space distribution of the  $L\sim L_\ast$ galaxies is a useful approximation to the mass distribution that fits the $\Lambda$CDM theory to the measured anisotropy spectrum of the Cosmic Microwave Background (Planck Collaboration et al. 2021 and references therein). Thus I take it that the reference 2MRS sample detectable out to redshift $z=0.02$ likely is a useful approximation to the large-scale mass distribution.  

\section{Distributions in Supergalactic Latitude}\label{sec:angularpositions}

This section deals with distributions of supergalactic latitudes as measures of  concentrations to the plane of the Local Supercluster. Several varieties of extragalactic objects are considered. 

\begin{figure}
\begin{center}
\includegraphics[angle=0,width=3.in]{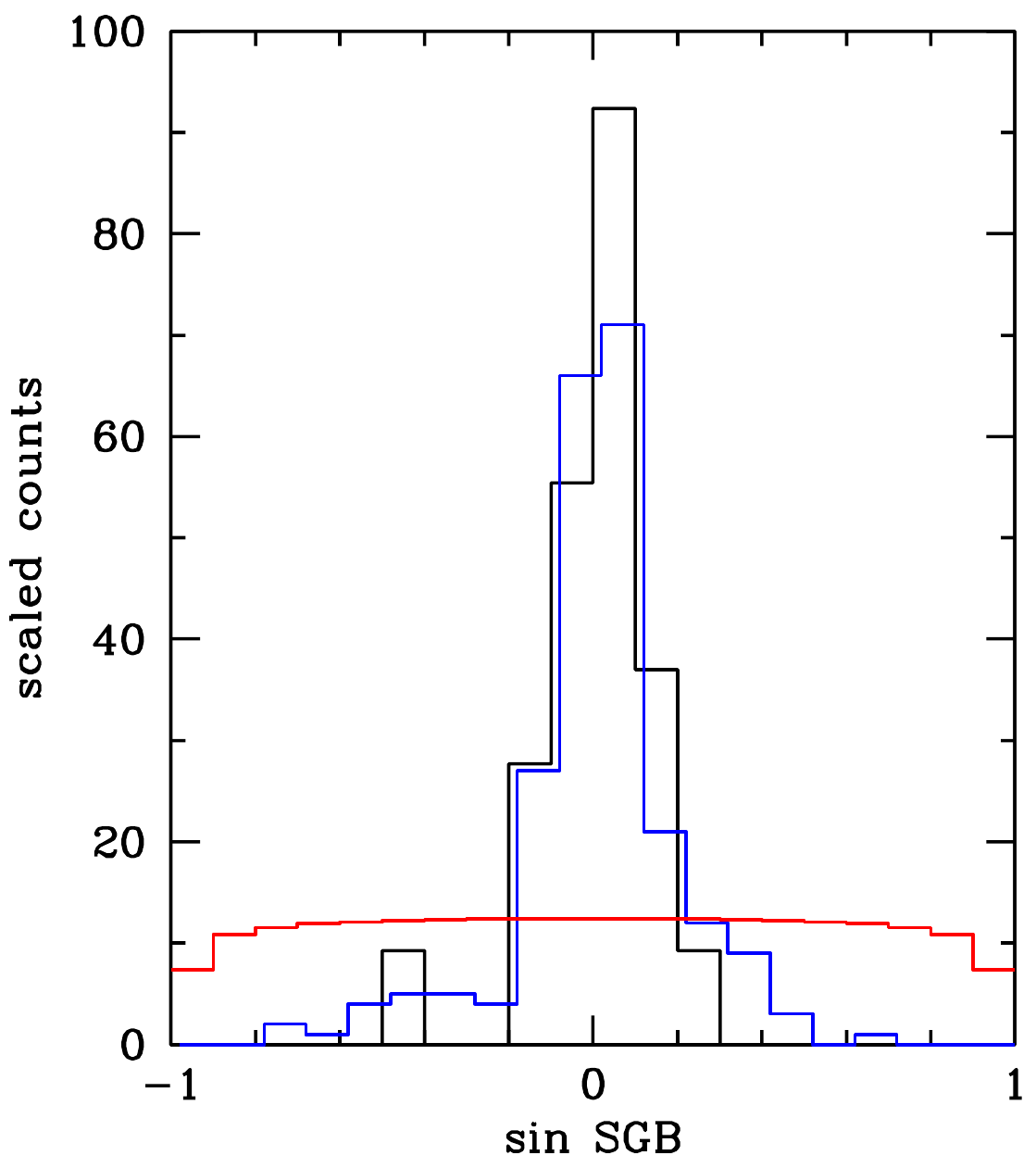} 
\caption{Distributions of supergalactic latitudes of the galaxies at distances of 2 to 6~Mpc, plotted in black for the 25 most luminous galaxies, plotted in blue and shifted slightly to the right for the 231 less luminous galaxies, and in red for the expected close to flat distribution of a fair sample of an isotropic random process truncated at galactic latitude $|b|=10^\circ$. The counts are scaled to a common area under each histogram.}\label{fig:LU}
\end{center}
\end{figure}

\subsection{Nearby Galaxies}

The nearby galaxies are known to be concentrated to the plane of the Local Supercluster. The data for the illustration in Figure~\ref{fig:LU} were drawn from the  Catalog and Atlas of the LV galaxies\footnote{Available at \url{https://www.sao.ru/lv/lvgdb/tables.php}. Angular positions and distances are in separate tables that do not include all the same galaxies. I have checked for matching galaxy names.} expanded from Karachentsev, Karachentseva, Huchtmeier, and Makarov (2004). I use the galaxies for which there are tabulated nonzero distance modulus uncertainties eDM$ < 0.434$, for fractional distance uncertainties less than 20\%. The histograms in Figure~\ref{fig:LU} show the counts of LV galaxies in the equal bins of $\sin{\rm SGB}$ indicated by the horizontal line segments, and at distances $2<r<6$~Mpc. The lower bound removes Local Group galaxies that are close to the plane of the Local Supercluster but might be at high supergalactic latitude. The upper bound is about at the edge of McCall's (2014) Local Sheet.

The histogram plotted as the black lines  in Figure~\ref{fig:LU} is the distribution of the 25 galaxies that are most luminous at absolute B magnitude. The vertical axis is scaled to get  the same area under this histogram as the distribution plotted in blue, and displaced slightly to the right, for the 231 less luminous galaxies. The vertical axis is the actual counts of these less luminous galaxies. The close to flat red histogram is the distribution expected of an isotropic random process cut off at $|b|=10^\circ$, again with the same area under the curve. It was computed from counts of positions uniformly placed at randomly at galactic latitude $|b|>10^\circ$, and again normalized to the same area under the  curve.

The prominent peaks in the distributions in Figure~\ref{fig:LU} illustrate the known tendency of nearby galaxies to be close to the plane of the Local Supercluster. The peaks would be even tighter if coordinates had been referred to the slightly tilted Local Sheet (McCall 2014). Within the modest precision afforded by just 25 galaxies in the more luminous sample the histograms of distributions of luminous and faint galaxies look to be essentially the same. This effect is discussed further in Section~\ref{sec:Challenges}. But the point of particular relevance for the discussion to follow is the close concentration of the nearby galaxies to the plane of the Local Supercluster.  

\begin{figure}
\begin{center}
\includegraphics[angle=0,width=3.25in]{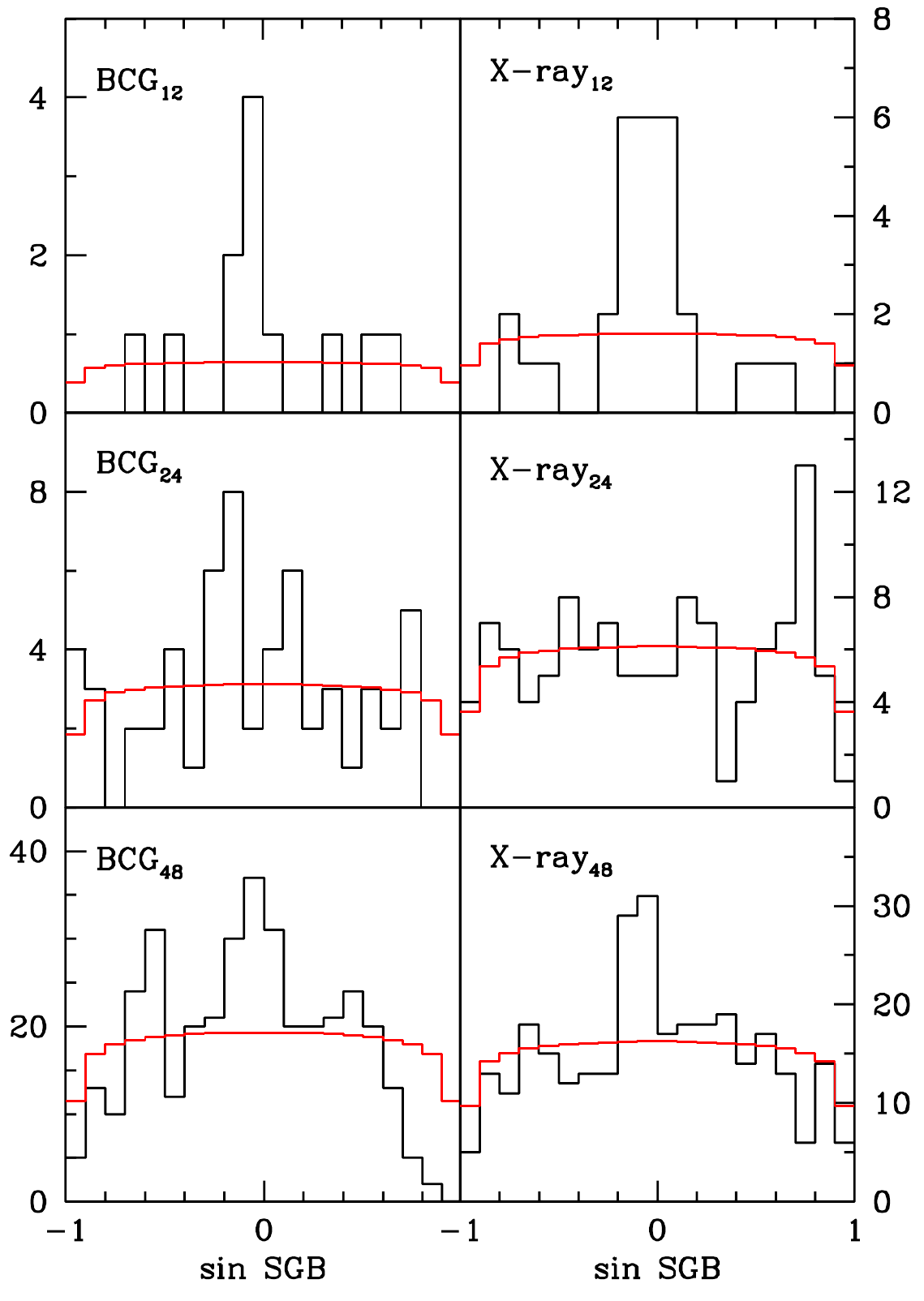} 
\caption{Distributions of supergalactic latitudes of clusters of galaxies identified by the Lauer et al (2014) compilation of redshifts of Brightest Cluster Galaxies, BCGs, and by the NASA HEASARC compilation of X-ray luminous clusters. The redshift bounds are indicated by the subscripts on the panel names: 0.01 to 0.02, 0.02 to 0.04, and 0.04 to 0.08.}  \label{histBCG}
\end{center}
\end{figure}

\subsection{Clusters of Galaxies}

The distributions of galaxies and clusters of galaxies at redshifts greater than $z=0.01$, distances $D \gap 45$~Mpc, are beyond a conventional size of the Local Supercluster, but they are in the range of distances at which Einasto et al. (1983), Shaver and Pierre (1989), and Shaver (1990, 1991) saw evidence of alignment to the plane of the Local Supercluster. The angular distributions of clusters of galaxies at these greater distances are illustrated by the two samples shown in Figure~\ref{histBCG}. The data on the left are from the Lauer, Postman, Strauss, Graves, and Chisari (2014) catalog of redshifts of the Brightest Cluster Galaxies, BCG, in the  catalogs of clusters compiled by Abell~(1958) and Abell, Corwin, and Olowin (1989). These data were downloaded from the SIMBAD database. The right-hand column shows distributions of X-ray sources identified as intracluster plasma. (These data were downloaded from the NASA HEASARC compilation.) The subscripts on the panel labels are reminders of the three ranges of redshift sampled: $0.01<z<0.02$, $0.02<z<0.04$, and $0.04<z<0.08$. As in Figure~\ref{fig:LU}, Figure~\ref{histBCG} shows histograms in equal bins of $\sin\hbox{SGB}$, and the near flat red curves are the mean distributions of an isotropic process truncated at $|\ell | = 10^\circ$ and normalized to the same area under the curve.

I use the full BCG catalog without bounds on the galaxy luminosity or the cluster richness. I remove from the sample of X-ray sources the faint tail of clusters with compiled measures of the X-ray luminosity less than $10^{42}$~erg~s$^{-1}$. This removes three clusters at $0.01<z<0.02$, leaving 30 X-ray clusters with 12 BCGs, for volume number densities $1.6\times 10^{-5}$~Mpc$^{-3}$ and $6\times 10^{-6}$~Mpc$^{-3}$ (from Eq.~[\ref{eq:densities}]). It removes 11 clusters at $0.02<z<0.04$, leaving 114 X-ray clusters with 58 BCGs. It leaves all 302 X-ray clusters at $0.04<z<0.08$ with 359 BCGs.

The distributions of sin~SGB in the two samples in the lowest redshift bin are quite similar but differ in number and detail, meaning Abell's visual selections and the X-ray detections  differently sample the varieties of groups and clusters of galaxies. As Einasto, Corwin, Huchra et al. (1983) saw, and B{\"o}hringer, Chon, \& Tr{\"u}mper (2021a) checked in their compilation of clusters detected by the ROSAT All Sky Survey, clusters of galaxies at redshifts up to $z=0.02$, within distance $\sim 85$~Mpc, are closely concentrated to low SGB, aligned with the plane of the Local Supercluster.

The peaks in the distributions of sin~SGB at greater distances, in the bottom two rows of panels in Figure~\ref{histBCG}, are not as prominent. It seems likely to be  significant, however, that in three of the four panels the peaks are at low SGB. This is discussed further in Section~\ref{sec:conclusions}. 

\begin{figure}
\begin{center}
\includegraphics[angle=0,width=1.75in]{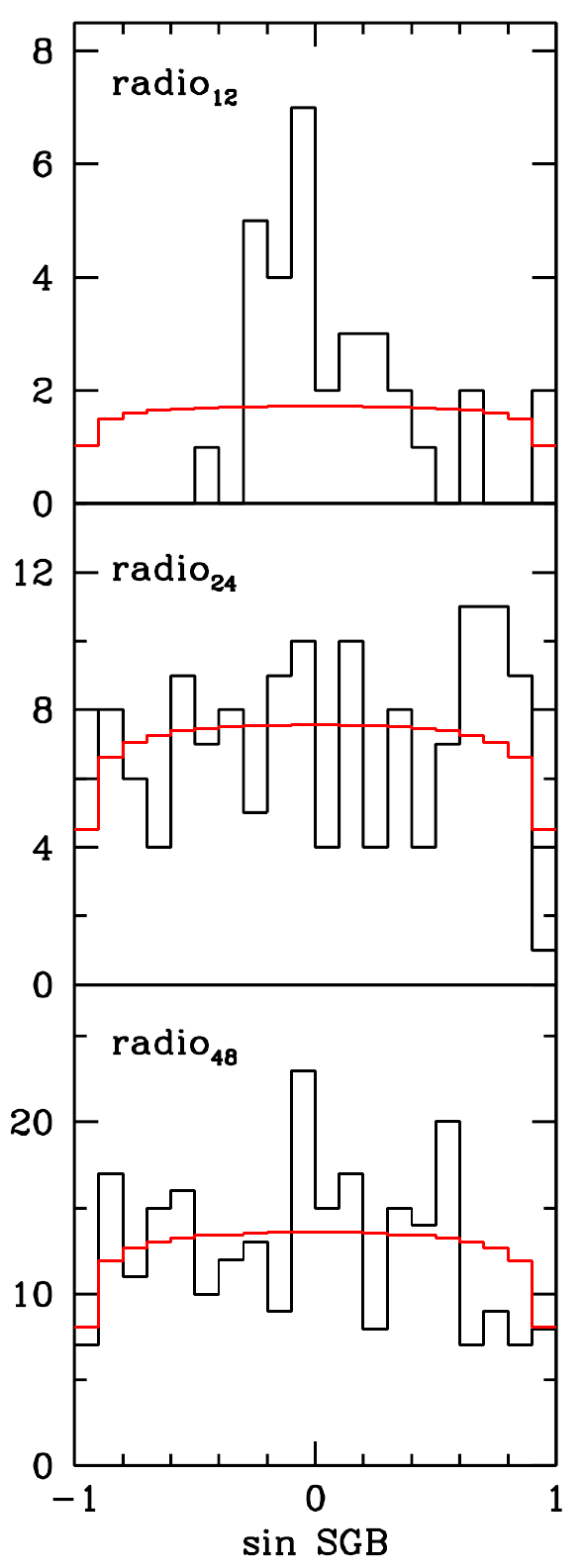} 
\caption{Distributions of supergalactic latitudes of radio-luminous galaxies, in the data compiled by van Velzen et al. (2012).}  \label{hist-radio}
\end{center}
\end{figure}

\subsection{Radio Galaxies}\label{sec:radio}

Shaver and Pierre (1989) showed evidence that radio galaxies at redshifts $z< 0.02$ tend to be close to the plane of the Local Supercluster. Figure~\ref{hist-radio} illustrates the present situation in the data drawn from the all-sky catalog of radio galaxies compiled by van Velzen, Falcke, Schellart, Nierstenh{\"o}fer, and Kampert (2012; the data were downloaded from the VizieR Online Data Catalog). As in Figure~\ref{histBCG}, the subscript on the panel name represents the redshift range. The samples used here are bounded by the Van Velzen et al. compilation of radio luminosities interpolated to 1.1 GHz, $\nu L_\nu>3\times 10^{39}~\hbox{ergs s}^{-1}$. This cutoff was chosen to make the counts of radio galaxies in Table~\ref{tab:samplenumbers} for the distributions in the three redshift bins shown in Figure~\ref{hist-radio} comparable to the counts of clusters represented by BGGs or intracluster plasma X-ray sources. 

Consistent with what Shaver and Pierre (1989, Fig. 5) saw, the distinct peak in the distribution of sin~SGB in the redshift range $0.01<z<0.02$, in the top panel in  Figure~\ref{hist-radio}, shows that radio galaxies at distances out to 85~Mpc are concentrated near the plane of the Local Supercluster, along with clusters of galaxies. Some of these radio galaxies are in clusters, and some surely are in rich groups; Centaurus~A is a nearby example not in the redshift range of the figure. The differences in details of 
the histograms in Panels BCG$_{12}$, X-ray$_{12}$, and radio$_{12}$ show that observations of radio galaxies are in effect yet another way to sample the varieties of groups and clusters. 

The distribution of supergalactic latitudes of radio galaxies at redshifts  $0.02<z<0.04$ is reasonably consistent with statistical isotropy. This can be taken to suggest that the position of the peak at low SGB in the more distant sample, $0.04<z<0.08$, likely is accidental. But arguing against this is the  coincidence that the position of this peak agrees with the peaks in three of the four distributions of clusters at redshifts greater than $z=0.02$ in Figure~\ref{histBCG}.

\begin{figure}
\begin{center}
\includegraphics[angle=0,width=2.75in]{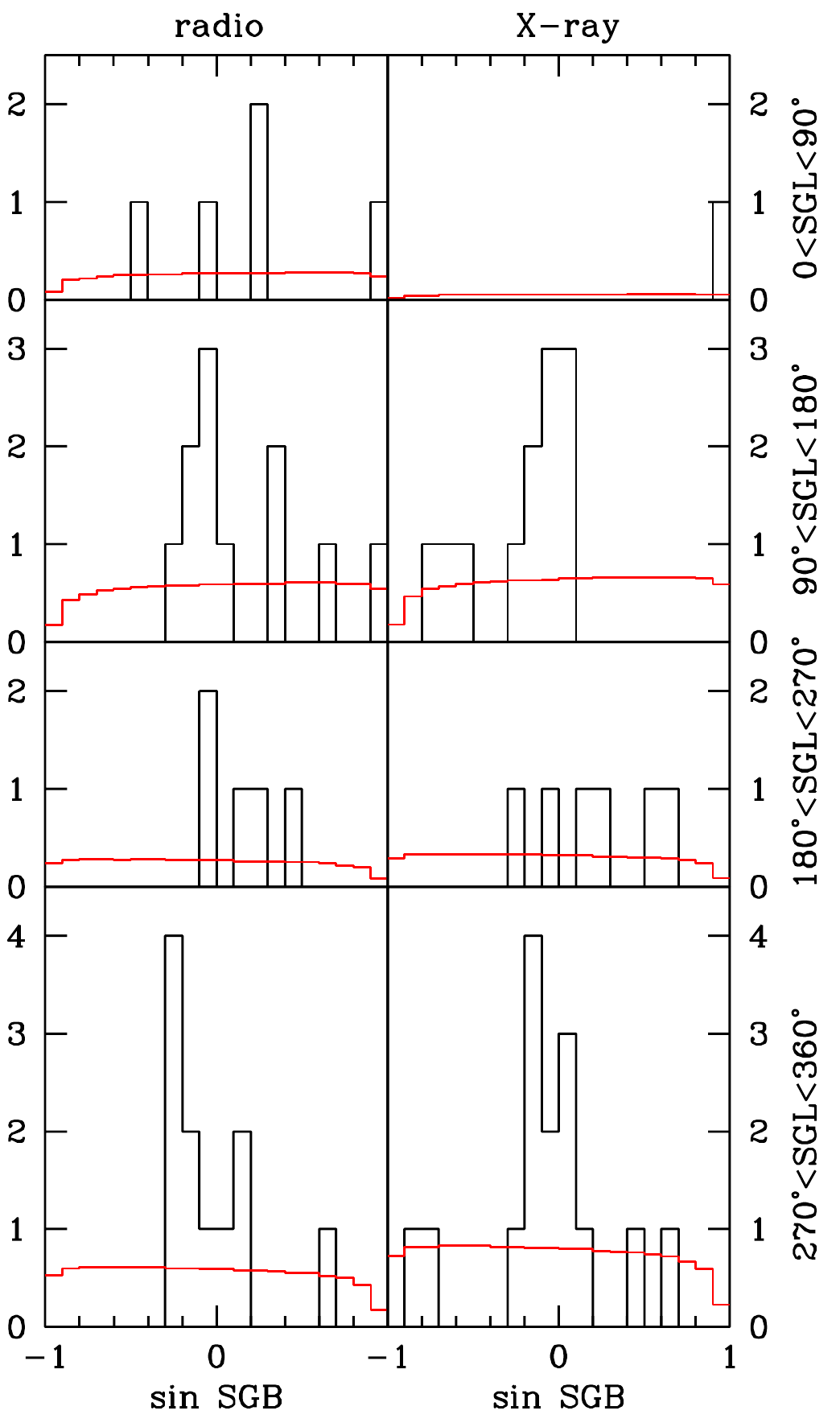} 
\caption{Angular distributions of radio galaxies and clusters of galaxies at $0.01<z<0.02$ in supergalactic latitude and quadrants of supergalactic longitude.}  \label{fig:quad}
\end{center}
\end{figure}

\subsection{Distributions in Supergalactic Longitude}

The four rows of histograms in Figure~\ref{fig:quad} show distributions of radio galaxies and clusters of galaxies in sin~SGB and in separate quadrants of supergalactic longitude SGL. This is a check that at $0.01<z<0.02$ the tendencies to low supergalactic latitudes of these objects are best described as clumpy distributions across the two dimensions projected on the plane of the Local Supercluster. 

If considered separately each of the distributions  in Figure~\ref{fig:quad} would not be likely to attract attention, but instead would bring to mind the general clumpy distributions of extragalactic objects. As already seen in the distributions summed over supergalactic latitude in Panel~X-ray$_{12}$ in Figure~\ref{histBCG} and Panel~radio$_{12}$ in Figure~\ref{hist-radio}, most of these objects are at low SGB. The point added in Figure~\ref{fig:quad} is that there are radio galaxies in comparable numbers in all four quadrants, and that while only one cluster is detected as an X-ray source in the first quadrant, and it is at high SGB, there are larger numbers of cluster X-ray sources in each of the other three histograms, and most are at low SGB. 

I conclude from Figure~\ref{fig:quad} that in a region around us that is 170~Mpc across the radio galaxies and clusters of galaxies are scattered around the plane of the Local Supercluster, and significantly closer to the plane than the size of this region. 

\begin{figure}
\begin{center}
\includegraphics[angle=0,width=2.75in]{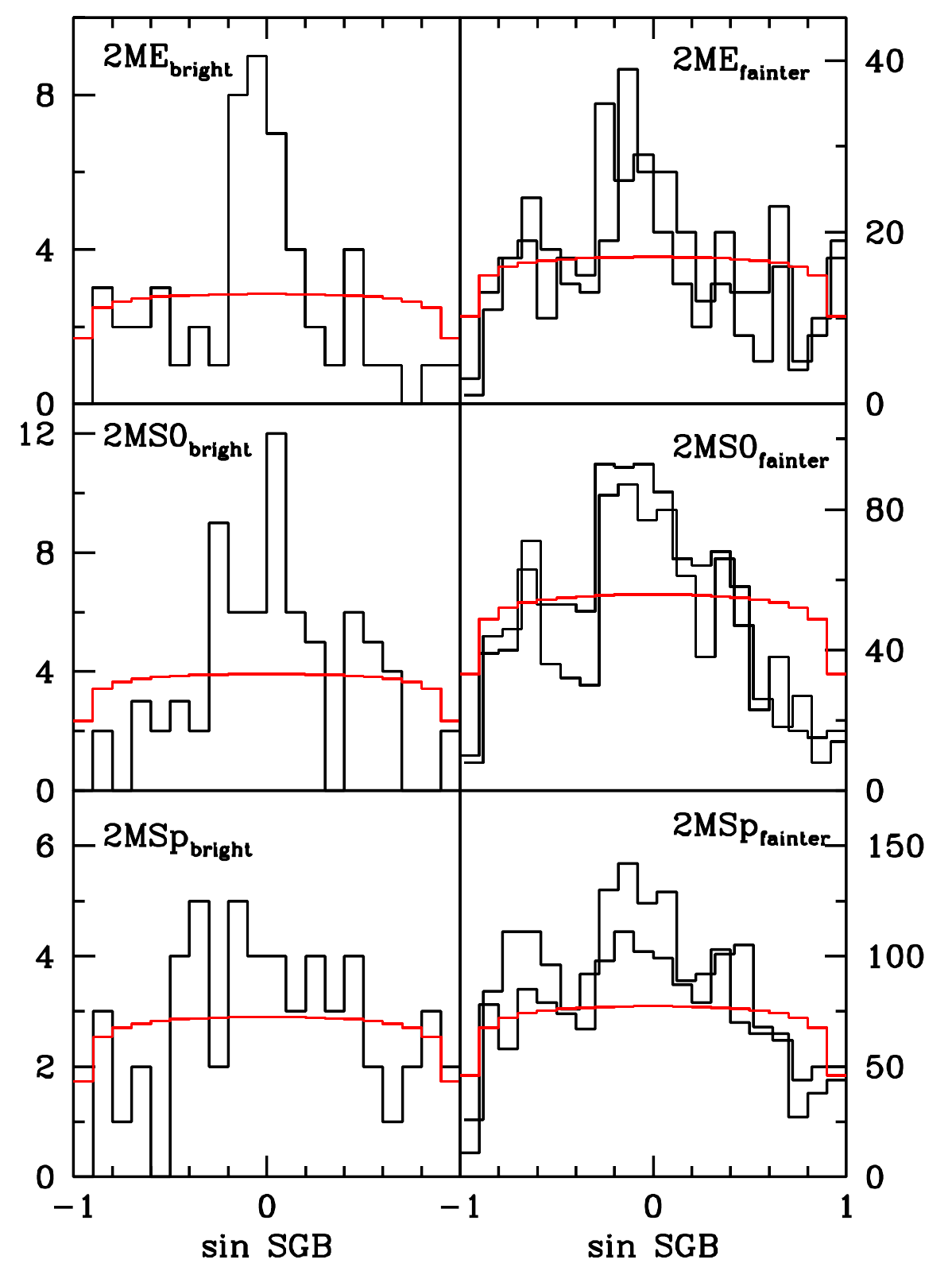} 
\caption{Distributions of elliptical, S0, and spiral  2MRS galaxies from the 2MASS survey at redshifts $0.01<z<0.02$ in three bins in absolute magnitude: most luminous on the left, intermediate luminosity on the right, and also on the right the least luminous of each morphological type shifted slightly to the right.}  \label{fig:ESp}
\end{center}
\end{figure}

\subsection{Distributions of Morphological Types}

It has long been known that galaxies that are exceptionally luminous at radio wavelengths tend to be in large elliptical or irregular galaxies in clusters and rich groups of galaxies. For example, in a pioneering paper on synchrotron radiation as the source of radio galaxy emission, Burbidge (1958a) commented on the radio galaxies in the Virgo and Perseus clusters and possibly in the Coma cluster; Einasto, Joeveer, and Saar (1980) presented maps showing that radio galaxies tend to be in or in the neighborhood of rich clusters of galaxies that are in superclusters; and in the presentation of the radio galaxy catalog used here Van Velzen et al. (2012) reported that ``the majority of the radio galaxies in our sample have massive, early-type hosts.''

To follow up on this I show in Figure~\ref{fig:ESp} distributions in sin~SGB of the early- and late-type galaxies at redshifts $0.01<z<0.02$ drawn from the 2MRS redshift catalog of 2MASS galaxies (Skrutskie et al. 2006; Huchra et al. 2012). At the absolute magnitude limits in the figure the apparent magnitude cutoff at redshift $z=0.02$ is $K_s=9.5$ in the histograms for the most luminous subsets in the left-hand panels. The intermediate and lowest luminosity limits in the distributions in the right-hand panels correspond to apparent magnitude bounds $9.5<K_s<10.9$ and $10.9<K_s<11.75$ at redshift $z=0.02$. The histograms for  the least luminous galaxies in the right-hand panels are shifted to larger sin~SGB by 0.02  to separate them from the distributions of the intermediate luminosity galaxies. The red near flat curves in the right-hand panels are normalized to the counts of the intermediate luminosities of each type.   

The morphological types are sorted by the Huchra et al. (2012) tabulation of the morphological parameter T: elliptical galaxies with T$\leq -5$ are in Panels 2ME; lenticular and S0 galaxies, here collectively termed S0, with $-5<$T$\leq 0$, are in Panels 2MS0; and spiral galaxies with $1\leq{\rm T}\leq 9$ are in Panels 2MSp. Galaxies with $T>9$ include various kinds of irregulars along with those that are difficult to classify or have not been visually examined. They are not entered in these distributions. In the most luminous, less luminous, and least luminous of the samples there are respectively 53, 319, and 325 ellipticals; 73, 1038, and 900 S0s; and 54, 1440, and 1782 spirals at $0.01<z<0.02$.

Consistent with the observed concentration of clusters of galaxies to low SGB, and the tendency of clusters to contain early-type galaxies, the distributions of the most luminous of the early-type galaxies in Panels~2ME$_{\rm bright}$ and~2MS0$_{\rm bright}$ in the left-hand side of Figure~\ref{fig:ESp} have prominent peaks at low SGB. In this redshift bin there are 126  luminous early type 2MRS galaxies, ellipticals plus S0s, considerably more than the 12 optically selected BCGs or the 30 clusters detected as  X-ray sources. That is, most of these luminous ellipticals and S0s are in groups rather than clusters. Perhaps the fractions of the most luminous ellipticals and S0s at high SGB are somewhat larger than the fractions of BCGs and radio galaxies, but the samples are too small to allow a clear case.

At the adopted luminosity cut, about two-thirds of the most luminous of the classified 2MRS galaxies are early types, with the rest spirals. The angular distribution of these luminous spirals in Panel~2MSp$_{\rm bright}$ has a modest peak at sin~SGB~$\sim -0.3$, but the number of spirals is too small to judge its likely significance. We can only conclude that, if a tendency to larger counts at lower SGB is present, then the excess of luminous spirals at low SGB is more modest than for ellipticals and S0s. 

The two disjoint lower luminosity bins contain totals of 2582 early types, ellipticals plus S0s, and 3222 spirals. The lower fraction of early types in the lower luminosity bins agrees with the known increasing ratio of early to late types with increasing luminosity. 

In all six histograms in the right-hand panels in Figure~\ref{fig:ESp} we see a reasonably clear indication of excess numbers over isotropy at low SGB. The anisotropy is least in the two disjoint spiral samples in Panel~2MSp$_{\rm fainter}$. Considering the small number of the most luminous spirals in the distribution in Panel~2MSp$_{\rm bright}$, there could be a fractional departure from isotropy consistent with what is seen in the lower luminosity late-type 2MRS galaxies. That is, the fractional anisotropies in sin~SGB favouring low supergalactic latitudes might be present in all three spiral samples, but the effect is smaller than in the six early type samples, and much smaller than the concentrations of radio galaxies and clusters of galaxies to low SGB. This refines what Shaver (1991) saw in the samples available then. 

\begin{figure}
\begin{center}
\includegraphics[angle=0,width=3.25in]{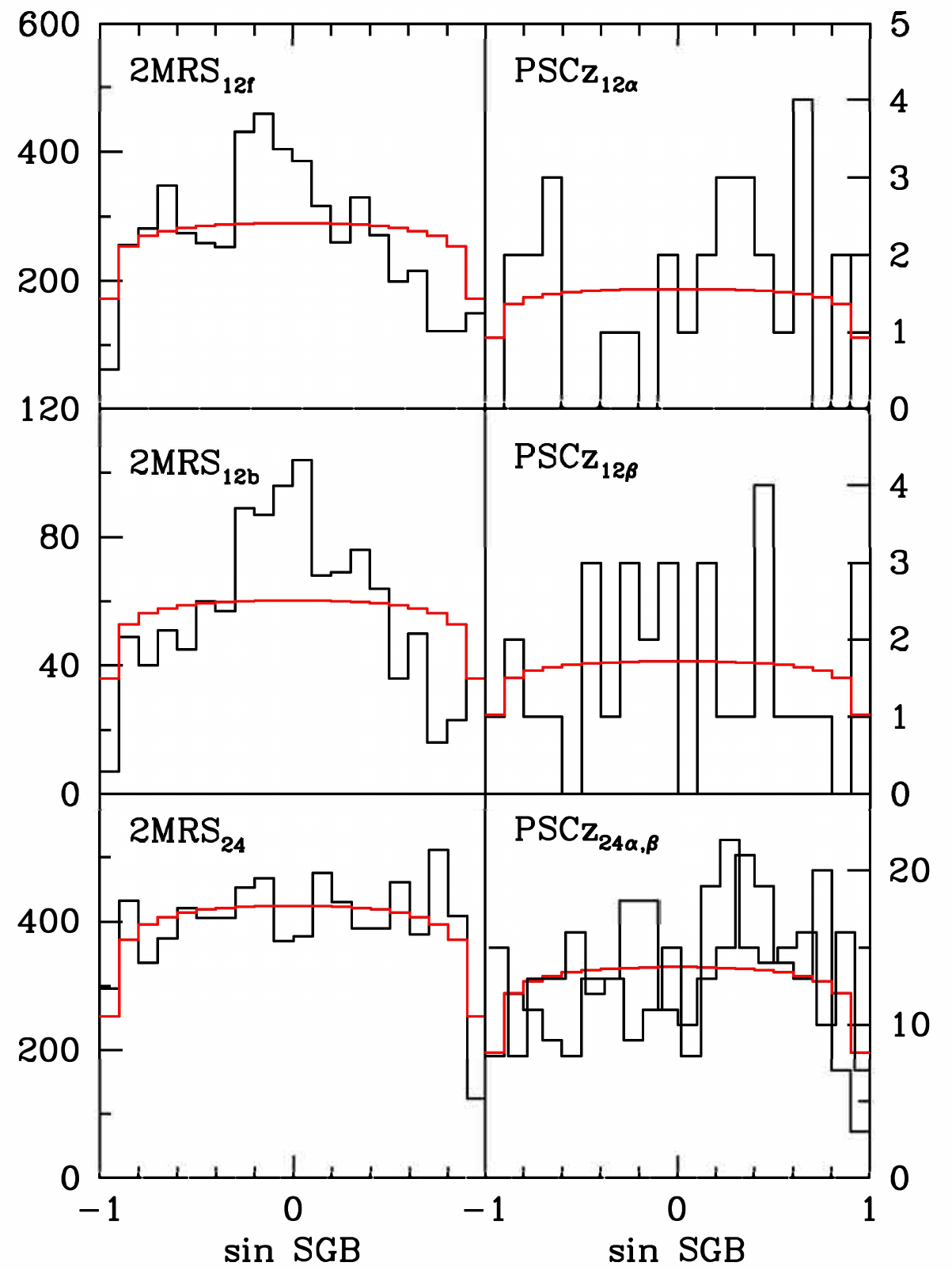} 
\caption{Comparisons of distributions of 2MRS (Huchra et al. 2012) and PSCz (Saunders et al. 2000) galaxies. The samples in panels~2MRS$_{\rm 12b}$ and 2MRS$_{24}$ are complete to the same limiting absolute magnitude. The sample 2MRS$_{\rm 12f}$ is complete to $z=0.02$ but does not include the more luminous 2MRS$_{\rm 12b}$ galaxies. Panels~PSCz$_{12\alpha,\beta}$ show angular distributions of the 29 galaxies at $z=0.01$ to $0.02$ that are most luminous at $60\mu$ and the next 32 most luminous. The histograms  in panels PSCz$_{24\alpha, \beta}$ are at the same two luminosity bounds as PSCz$_{12\alpha, \beta}$ respectively.}\label{histPSCz}
\end{center}
\end{figure}

\subsection{Luminous PSCz Galaxies }

Figure~\ref{histPSCz} compares the distributions of supergalactic latitudes of the 2MRS galaxies selected at the 2MASS $2.2\mu$ wavelength band and the PSCz galaxies selected for this figure at $60\mu$. The former includes all morphologies in the Huchra et al. (2012) catalog, and all the objects without an estimate of morphology but with measured redshifts and apparent magnitudes in the wanted ranges, because these data argue for detections of galaxies of some sort. It does not include galaxies with listed redshifts but without entries for the uncertainty in the redshift, because the redshifts might be seriously uncertain. The PSCz samples in the right-hand side of Figure~\ref{histPSCz} are from the galaxy redshift catalog that Saunders, Sutherland, Maddox, et al. (2000) drew from the IRAS (infrared astronomical satellite) sky survey at wavelengths from 12 to $100\mu$. This valuable compilation is sampled in a close to uniform way across a large part of the sky. (The data used here were downloaded from the NASA HEASARC IRASPSCZ catalog, class GALAXY.)

Luminous Infrared Galaxies, LIRGs (Magnelli, Elbaz, Chary, et al. 2011 and references therein), are a subset of the PSCz galaxies. I have not attempted to separate LIRGs from the more luminous ULIRGs and the less luminous IRAS galaxies that are not considered LIRGs; the name is used here as a convenient abbreviation for the most luminous of the IRAS PSCz objects to be considered here, with space number density comparable to that of clusters of galaxies.

Panel 2MRS$_{\rm 12f}$ in Figure~\ref{histPSCz} shows the distribution in SGB of  the galaxies that are luminous enough to be detected at $z=0.02$, at limiting apparent magnitude $K_s = 11.75$, but not luminous enough to be detected at $z=0.04$. Panel~2MRS$_{\rm 12b}$ shows the distribution of the disjoint sample of PSCz galaxies in the lower redshift bin that are luminous enough to be detected at $z=0.04$. This can be compared to the distribution in Panel~2MRS$_{24}$ at $z=0.02$ to 0.04 and the same luminosity bound as for 2MRS$_{\rm 12b}$.  

The 5393 2MRS galaxies in the lower luminosity sample in Panel 2MRS$_{\rm 12f}$ amount to space number density 0.003~Mpc$^{-3}$. This is comparable to the density of ordinary $L\sim L_\ast$ galaxies, meaning the space distribution of these galaxies likely is a useful approximation to the spatial distribution of mass (as argued in Sec.~\ref{sec:data}). The angular distribution in this sample has a distinct peak at low SGB, but a considerable fraction of these galaxies is at higher supergalactic latitudes. The distribution in the disjoint sample of 1123 more luminous PSCz galaxies in Panel~2MRS$_{\rm 12b}$ is quite similar. As expected, it also is reasonably similar to the distributions of 2MRS galaxies separated by morphological types, in the right-hand panels in Figure~\ref{fig:ESp}. But the distributions in Panels~2MRS$_{\rm 12b,f}$ differ from the tighter concentrations of the most luminous of the early types in the top two left-hand panels in Figure~\ref{fig:ESp}. 

The 2MRS galaxies in panels 2MRS$_{\rm 12b}$ and 2MRS$_{24}$ are luminous enough to be detected at redshift $z=0.04$ at the $2.2\mu$ limiting magnitude $K_s\leq 11.75$. The common absolute magnitude limit allows comparison of distributions at about the same number density in the two redshift bins (at counts 1123 and 7908, the number density is about $6\times 10^{-4}$~Mpc$^{-3}$). The difference is pronounced: the counts at $0.02<z<0.04$ look like an isotropic random process. The distribution at $0.04<z<0.08$ is not shown; it too looks like an isotropic random process.

For the present purpose the point of particular interest in Fgure~\ref{histPSCz} is that it completes the illustrations of Shaver's effect. The angular distribution of the 29 most luminous PSCz galaxies in Panel~PSCz$_{\rm 12\alpha}$ and, in the check of reproducibility in Panel~PSCz$_{\rm 12\beta}$, the  distribution of the 32 next most luminous PSCz galaxies, show no indication of a correlation or anticorrelation with angular position relative to the plane of the Local Supercluster. The most luminous spirals in Panel~2MSp$_{\rm bright}$ in Figure~\ref{fig:ESp} also are not more or less numerous than average near the plane of the Local Supercluster. These examples are to be compared to comparable numbers of radio galaxies in Panel~radio$_{12}$ in Figure~\ref{hist-radio} and luminous ellipticals in Panel~2ME$_{\rm bright}$ in Figure~\ref{fig:ESp}, which are tightly concentrated to low SGB, along with clusters of galaxies. 

We should pause to take note of the low counts of 2MRS galaxies near the north supergalactic pole in Panels 2MRS$_{\rm 12b}$ and 2MRS$_{\rm 12f}$, and in most of the panels in Figure~\ref{fig:ESp}. This departure from isotropy may or may not be related to the Local Supercluster. It should be noted also that the distribution of PSCz galaxies at a significantly lower luminosity cutoff, not shown, is peaked toward low SGB. This must be expected because at lower infrared luminosities the sample must include 2MRS galaxies and optically selected galaxies that tend to be at low SGB.

\subsection{A Summary Sky Map}

%\begin{landscape}
\begin{figure*}
\begin{center}
\includegraphics[angle=0,width=6.5in]{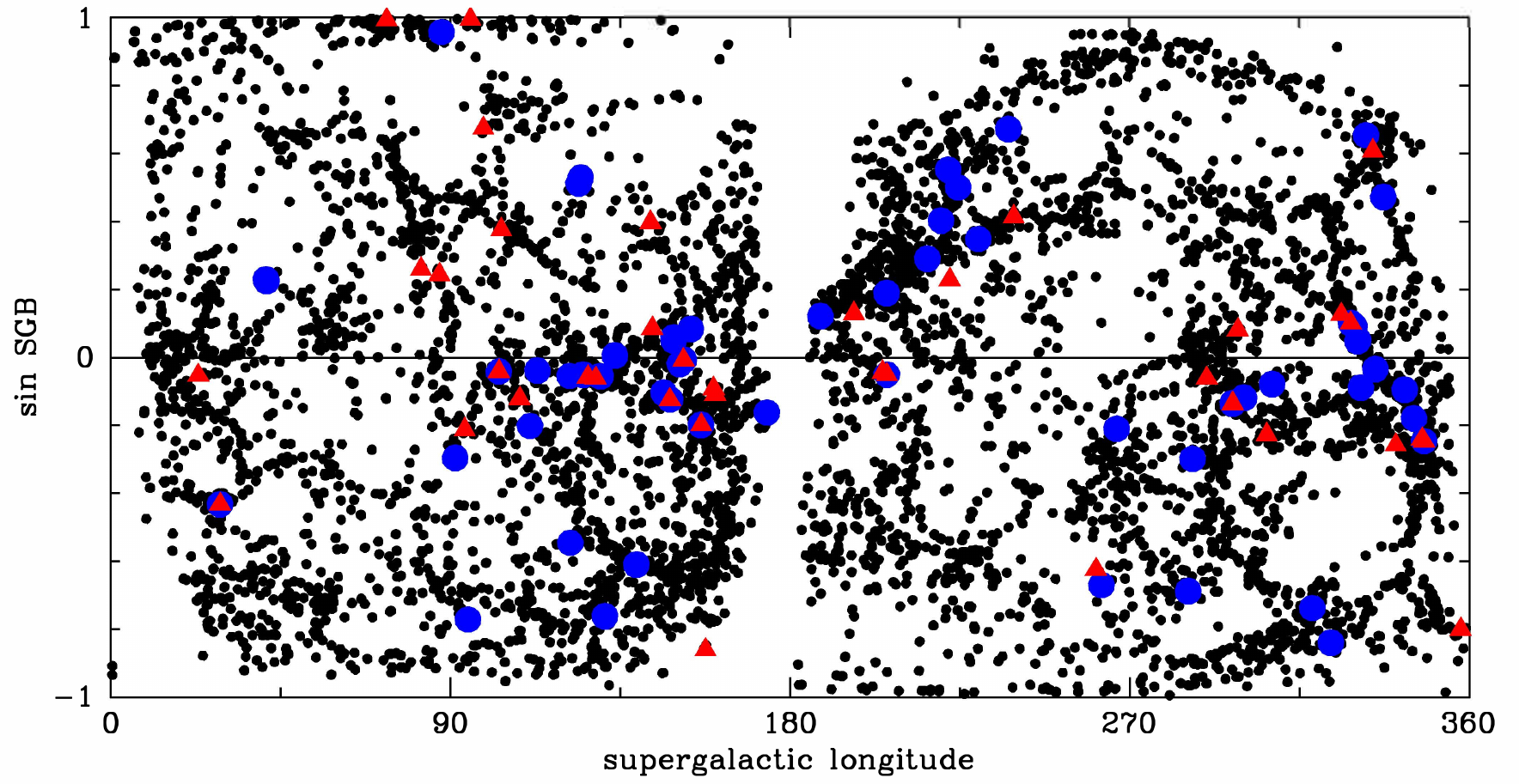} 
\caption{Map of extragalactic objects at $z<0.02$. The 2MRS galaxy positions are plotted as small black circles; radio galaxies are shown as larger red triangles; and clusters of galaxies as still larger blue circles.}\label{skyfig}
\end{center}
\end{figure*}
%\end{landscape}

The equal area map in Figure~\ref{skyfig} illustrates the distributions of galaxies and clusters of galaxies at redshift $z<0.02$, including those at low galactic latitudes. The plane of the Local Supercluster is close to the central horizontal line. The plane of the Milky Way runs close to the supergalactic pole ${\rm SGB} = -90^\circ$ in the left-hand side of the figure, then through SGB = 0 at supergalactic longitude near $180^\circ$, then up to the pole ${\rm SGB} = +90^\circ$ in the right-hand side. Obscuration by interstellar dust causes the empty regions at low galactic latitudes. This is taken into account in the previous figures by the truncation at galactic latitude $|b|=10^\circ$ of the samples and the red nearly flat curves showing  the expected distribution of an isotropic random process. 

The small black circles in the figure mark the positions of 2MRS galaxies that are luminous enough to be detected at $z=0.02$. This sample from the 2MASS survey is taken to be a good approximation to the distribution of $L\sim L_\ast$ galaxies, and a useful approximation to the mass distribution. The large blue circles show positions of the BGC and X-ray markers of clusters. Duplicates appear as overprinted blue circles. The red triangles show the positions of radio galaxies. A radio galaxy in a cluster appears as a triangle on a large circle. There are many examples of this in the figure, there are many clusters without a tabulated radio galaxy, and there are many radio galaxies not in tabulated clusters. 

Panels~2MRS$_{\rm 12b}$ and~2MRS$_{\rm 12f}$ in Figure~\ref{histPSCz} show counts of the galaxies represented as small black circles in Figure~\ref{skyfig} summed over supergalactic longitude. The counts in excess of isotropy at low supergalactic latitudes are seen in the two disjoint samples, and the anisotropy is in the direction of the excess counts of clusters and radio galaxies at low SGB. It is difficult to make all this out in the map in Figure~\ref{skyfig}. We see that the alignment to the plane of the Local Supercluster is not sharply defined; it is a statistical effect in the clumpy distribution of galaxies, but convincingly established by the reproducibility of the effect in the distributions in SGB reviewed in this section. 

\begin{figure}
\begin{center}
\includegraphics[angle=0,width=2.25in]{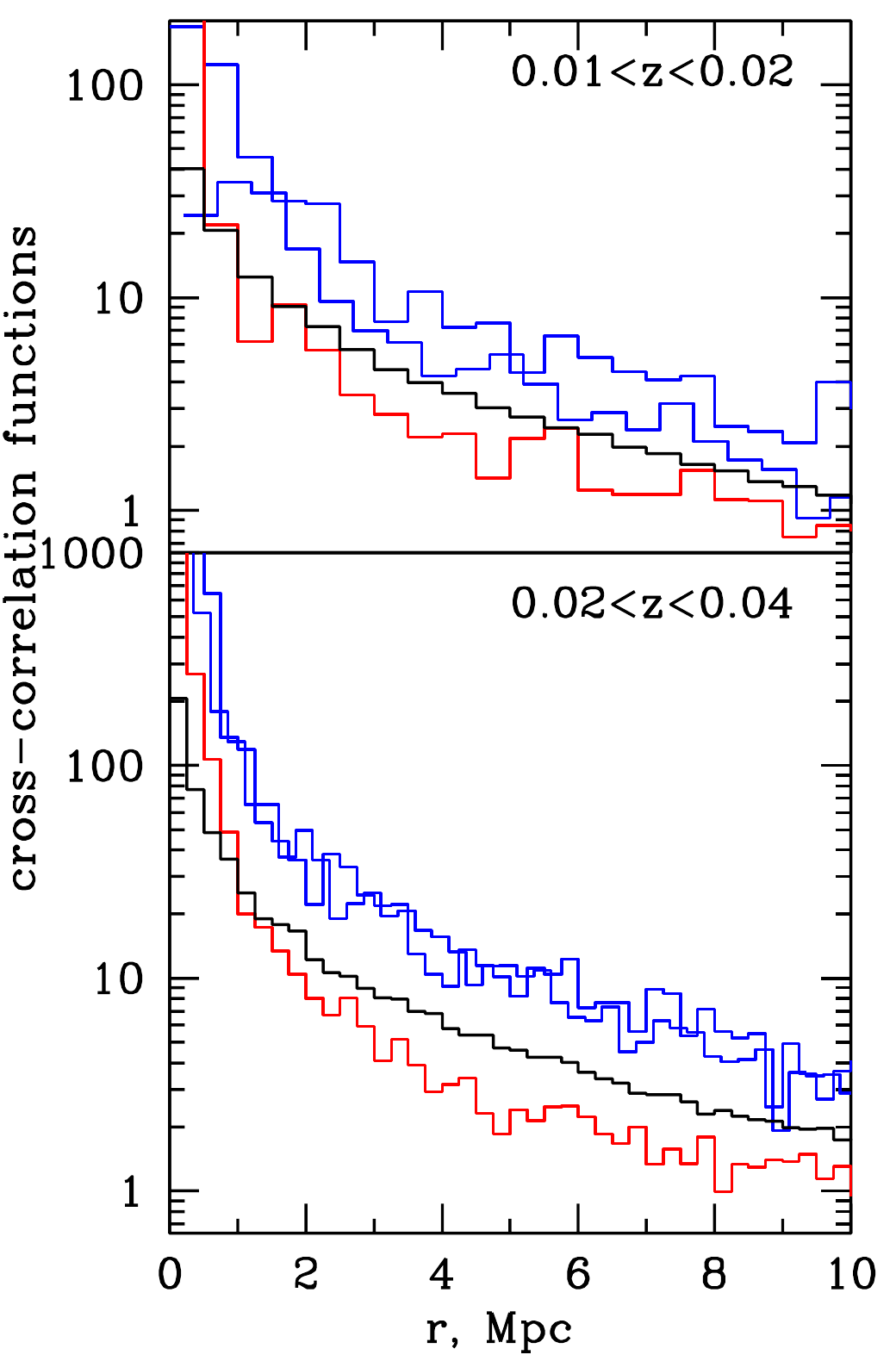} 
\caption{Cross-correlations of the positions of 2MRS galaxies with LIRGs in the lower red line, and with clusters of galaxies in the upper two blue lines. The middle black line is the autocorrelation function of the 2MRS galaxies.}  \label{corrlog}
\end{center}
\end{figure}

\begin{figure}
\begin{center}
\includegraphics[angle=0,width=2.in]{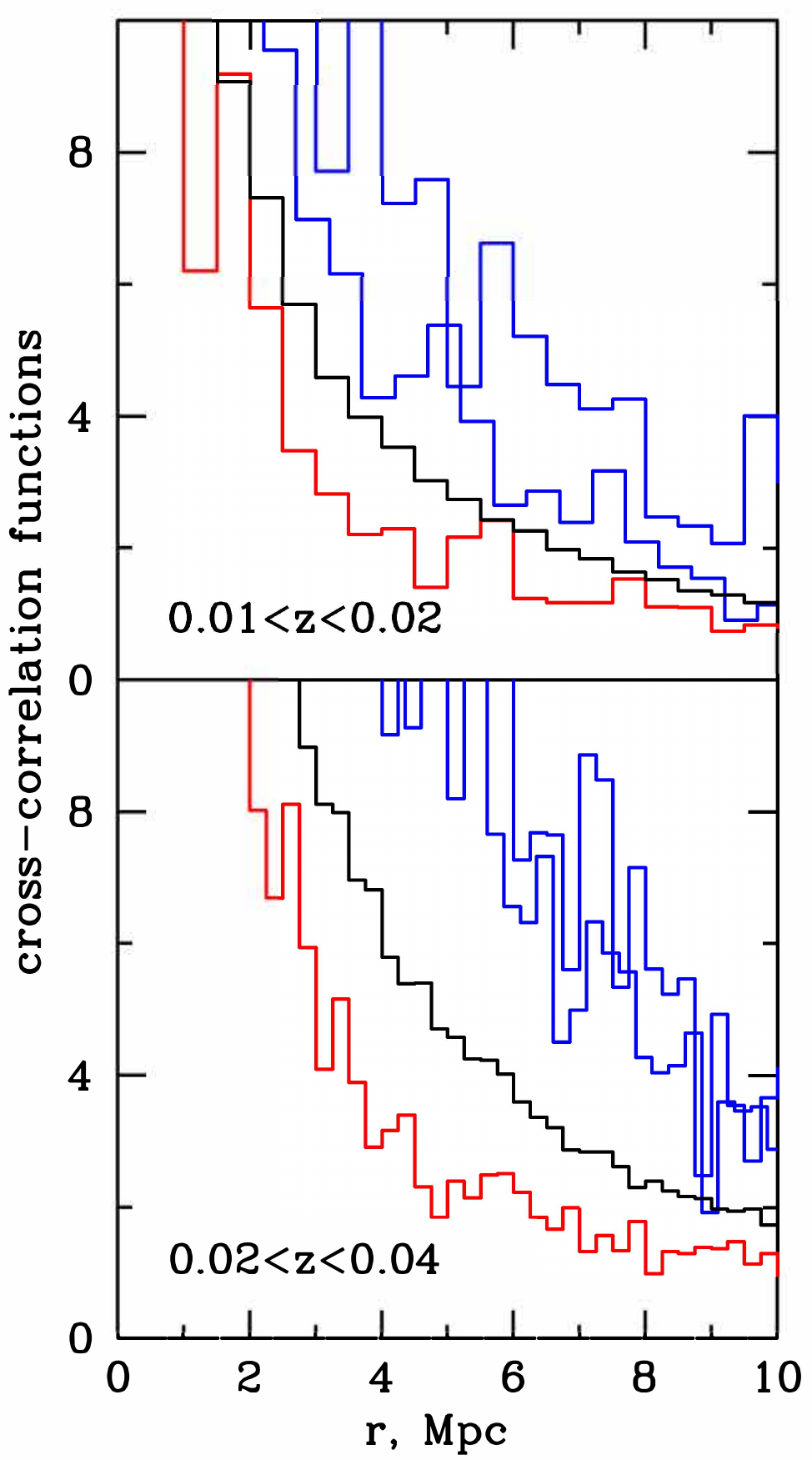} 
\caption{The same as in Figure~\ref{corrlog}, but plotted in linear coordinates.}  \label{corrlinear}
\end{center}
\end{figure}

\section{Position Correlation Functions}\label{sec:correlations}
 
 Some indication of the meaning of the differences in the distributions of different types of extragalactic objects relative to the Local Supercuster might be found in the cross-correlations of positions of these objects. Considered here are extreme examples that illustrate Shaver's effect: clusters of galaxies and luminous PSCz galaxies along with the numerous 2MRS galaxies that I take to approximate the mass distribution.
 
The two-point correlation functions are estimated in redshift space, where distance is proportional to redshift. Peculiar motions tend to smooth the two-point functions relative to real space, perhaps a larger effect for 2MRS galaxies closer to BCGs. Estimates of the two-point correlation as a function of orientation in redshift space, along the lines introduced in Davis and Peebles (1983) and Kaiser (1987), would enable separation of the effect of peculiar motions from the position space correlation, but the cluster and PSCz samples are too small for this. 
 
We have only modest constraints on the cross-correlation of cluster and LIRG positions at $0.02<z<0.04$. There is one BCG--LIRG pair separated in redshift space by less than 5~Mpc. A large number of points placed uniformly at random in the bounds of galactic latitude and redshift indicates that 1.0 pairs would be expected if positions were independent and random. At separation $5<r<10$~Mpc there are 18 pairs in the data, and 6.8 are expected if positions are independent and random. And there are 69 BCG--LIRG pairs at $10<r<20$~Mpc compared to 50 expected. Since the positions of these objects are not random we can only conclude from these numbers that there is not a strong tendency for LIRGs to be near clusters of galaxies, or to avoid them.

The larger number of 2MRS galaxies from the 2MASS survey offers better estimates of the cross-correlations of positions of the 2MRS sample with LIRGs and clusters. To begin, let us review the sample sizes. At $0.01<z<0.02$ there are 12 BCGs, 30 X-ray clusters, the sum of 62 LIRGs in Panels~PSCz$_{\rm 12\alpha}$ and PSCz$_{\rm 12\beta}$ in Figure~\ref{histPSCz}, and 6515 2MRS galaxies that are luminous enough to be detected at $z<0.02$. At $0.02<z<0.04$ there are 58 BCGs, 114 X-ray clusters, the sum of 523 LIRGs in Panels~PSCz$_{\rm 12\alpha,\beta}$, and 7908 2MRS galaxies that are luminous enough to be detected at $z=0.04$.  We must expect serious  shot noise from the small numbers of clusters and LIRGs, but the results are interesting.  

The cross-correlations are estimated as follows. In a sample of $N_{\rm S}$ objects let the count of objects in distance bin $n$ in redshift space from any of the 2MRS galaxies be Count$_{{\rm 2M,S},n}$. Place $N_{\rm R}$ points uniformly at random in the bounds of redshift and galactic latitude, and let Count$_{{\rm 2M,R},n}$ be the number of randomly placed points in the distance bin n from any 2MRS galaxy. If $N_{\rm R}$ is large enough the probability that a point placed uniformly at random in the bounds of the region falls in the distance bin $n$ from any of the 2MRS  galaxies in the region is Count$_{{\rm 2M,R},n}/N_{\rm R}$. This multiplied by the number $N_{\rm S}$ of sample objects is the expected count of neighbors at distance bin $n$ from any 2MRS galaxy, if the sample objects are uniformly distributed in the region.  This divided into the observed count gives an estimate of the cross-correlation between positions of 2MRS galaxies and the $N_{\rm S}$ objects in  distance bin $n$:
\beq
\xi_{\rm 2M,S}(n) = \frac{ {\rm Count}_{{\rm 2M,S},n}} {{\rm Count}_{{\rm 2M,R},n}}
		\frac{N_{\rm R}}{N_{\rm S}} - 1.  \label{eq:xi_2M-S}
\eeq
The random counts are computed separately for the two redshift bins.\footnote{One could instead measure the counts of random points in the distance bin $n$ from any of the sample points. Perhaps this or some other approach is a better way. I have not checked.}

Figures~\ref{corrlog} and~\ref{corrlinear} show the results of applying equation~(\ref{eq:xi_2M-S}) to the samples in two redshift bins, plotted in logarithmic and linear coordinates. The horizontal line segments in the histograms mark the intervals in separation $r$ within which equation~(\ref{eq:xi_2M-S}) is applied. The interval in $r$ for the smaller redshift bin is $\Delta r = 0.5$~Mpc. The larger numbers of clusters and LIRGs in the larger redshift bin allow a smaller interval, $\Delta r = 0.25$~Mpc. 

The cross-correlations of positions of 2MRS galaxies with the two cluster samples, BCGs and X-ray detections, are plotted as the uppermost two blue lines in both figures. The line for X-ray clusters is shifted to the right by 0.2~Mpc from the line for BCGs in the upper panel, and by 0.1~Mpc in the lower panel. In the redshift bin $z=0.01$ to~0.02 the two cluster-2MRS functions are systematically different, perhaps a simple result of the small sample sizes, 12 BCGs and 30 X-ray clusters. The samples are larger in the $0.02<z<0.04$ redshift bin, 58 and 114, and here the functions for the two cluster samples are tolerably consistent. 

The 2MRS position autocorrelation function plotted as the black centermost line in the two figures is estimated by application of equation~(\ref{eq:xi_2M-S}) with the number $N_{\rm 2MRS}$ of 2MRS galaxies treated as the sample:
\beq
\xi_{2M,2M}(n) = \frac{2\times{\rm Count}_{\rm 2M,2M,n}}
	{{\rm Count}_{\rm 2M,R,n}}
		\frac{N_{\rm R}}{N_{\rm 2MRS}} - 1.  \label{eq:xi_2M-2M}
\eeq
The factor of two is required if each distinct 2MRS pair is counted once. 

In both redshift bins the two 2MRS--cluster cross-correlation functions are larger than the 2MRS galaxy autocorrelation function. That is, the broad-spread concentration of mass around a typical cluster is greater than the mean concentration of mass around a typical galaxy. This was already seen in the angular cross-correlation of Abell clusters with Lick galaxy counts (Seldner and Peebles 1977). 

The lowest line in both figures, plotted in red, is the 2MRS-LIRG position cross-correlation function. At separations $\lap 0.5$~Mpc this function is much larger than the 2MRS autocorrelation function, and perhaps comparable to the 2MRS--cluster functions, though this depends on the correction for galaxy velocity dispersions. On this relatively small scale 2MRS galaxies are concentrated around LIRGs far  more than 2MRS galaxies are concentrated around 2MRS galaxies. At separation $r\gap 2$~Mpc, where the correction for peculiar velocities is less problematic, the mean 2MRS density around a LIRG is less than the density around a 2MRS galaxy, and well below the density around a cluster. This is consistent with other ways to establish that LIRGs tend to appear in dense but less extended concentrations of galaxies, which is to say in groups of galaxies (Tekola, V{\"a}is{\"a}nen, and Berlind 2012). Perhaps less expected is that at separations $5\lap r\lap10$~Mpc the LIRG-2MRS cross-correlation function is smaller than the 2MRS-2MRS function, and the  LIRG-2MRS function is positive but close to zero. On these scales the mean  density around a LIRG is smaller than the mean density around a typical 2MRS galaxy, and not much greater than the cosmic mean density.

\section{Discussion}\label{sec:conclusions}

I offer summary thoughts about observational issues that might be relevant for comparisons to numerical and analytic analyses of the formation of configurations similar to the extended Local Supercluster, and conclude with brief speculations.

\subsection{Observations}

It is well and clearly established that we are in a region at least 170~Mpc across in which most of the clusters of galaxies and radio galaxies, along with the most luminous early-type galaxies, are close to a common plane, the Local Supercluster. This is seen in the histograms in the top panels in Figures~\ref{histBCG} and \ref{hist-radio}, and in the top two left-hand panels in Figure~\ref{fig:ESp}. Since these objects all prefer dense regions the samples are related, but the differences among the histograms are to be expected because they are different ways to look at the nature of large-scale structure.

In contrast to this situation, the spirals at $z<0.02$ that are as luminous at 2.2~$\mu$ as the most luminous ellipticals are not noticeably concentrated to this plane, and the same is observed of the galaxies that are most luminous  at $60~\mu$ (as seen in Panel 2MSp$_{\rm bright}$ in Fig.\ref{fig:ESp} and Panels~PSCz$_{\rm 12\alpha,\beta}$ in Fig.~\ref{histPSCz}). These results generalize Shaver's (1991) effect: different kinds of extragalactic objects have different degrees of alignment to the plane of the Local Supercluster.

The Local Universe catalog\footnote{Available at \url {http://edd.ifa.hawaii.edu} as ``Local Universe (LU)'' in the Extragalactic Distance Database maintained by Brent Tully at the University of Hawaii and colleagues.} lists 31 galaxies at distances less than 10~Mpc with K-band luminosities greater than $10^{10}$. The large angular sizes of these galaxies require special treatment in the 2MASS survey, but it seems quite likely that similar galaxies at greater distances are common 2MRS galaxies. Images of the large ones in the Local Universe catalog can be found on the web; most are the usual beautiful spirals, some are ellipticals, and a few look odd to my untrained eye. The local number density of these 31 galaxies is 0.007~Mpc$^{-3}$. This is similar to the density of the 6515 2MRS galaxies detectable out to $z=0.02$, $n=0.003$~Mpc$^{-3}$, and it it is not far from the mean number density of the $L\sim L_\ast$ galaxies in an optically selected catalog. The space distribution of optically selected galaxies is a useful approximation to the mass distribution that fits the measured anisotropy of the thermal background radiation, and it is reasonable to assume the 2MRS galaxies similarly approximate the mass. This would mean the excess counts of 2MRS galaxies at low supergalactic latitudes (as in Panel 2MSp$_{\rm fainter}$ in Fig.\ref{fig:ESp}) signify a mass distribution at $z<0.02$ that  at low SGB is systematically and exceptionally larger than the cosmic mean. This would be consistent with the expected biased structure formation that accounts for the tight concentration of clusters to low SGB.

There is evidence of the extension of the alignment phenomenon beyond redshift $z=0.02$ from the coincidence of the well-established peaks in the distribution of sin~SGB at low SGB at $z<0.02$ with the peaks in the angular distributions of clusters and radio galaxies in Panels~BCG$_{24}$,  BCG$_{48}$, X-ray$_{48}$, and radio$_{48}$. We must bear in mind the absence of an interesting feature at low SGB in the distributions in Panels~X-ray$_{24}$, radio$_{24}$, and 2MRS$_{24}$. But the coincidences of the peaks in four of the seven distributions at higher redshifts makes an arguable case that the alignment phenomenon extends to redshift $z\sim 0.08$. This spans a region some 600~Mpc across. The length scale is remarkable; the evidence merits attention. 
 
Because we happen to be close to the plane of the Local Supercluster its presence is manifest in the spread of galaxies across the sky.  Surely there are similar configurations that are not so readily seen. An example of proposals of linear arrangements of galaxies on smaller on scales is the evidence of a planar alignment of large and dwarf galaxies around M\,101 extending to about 3~Mpc (M{\"u}ller, Scalera, Binggeli, and Jerjen 2017). On the larger scales considered in this paper, spreading over several hundred megaparsecs, one might think first of the Pisces-Perseus supercluster (Giovanelli, Haynes, and Chincarini 1986) and the Great Wall (Geller and Huchra 1989). Many similar more distant configurations are seen for example in the 6dF Galaxy Survey (Jones, Read, Saunders,  et al. 2009,  Figs. 8 and 9). B{\"o}hringer, Chon, and Tr{\"u}mper (2021a,b) present a program of investigation of these systems in X-ray detections. 

The maps of the distributions of galaxies of different morphological types in the Pisces-Perseus supercluster show that the distinct concentrations of early types along the ridge of this nearly linear configuration are less distinct among later types (Giovanelli, Haynes, and Chincarini 1986). This example of the morphology-density relation (Dressler 1997 and earlier references therein) also is seen in Figure~\ref{fig:ESp}: earlier-type galaxies are more strongly concentrated to the plane of the Local Supercluster. Open issues are whether radio galaxies and clusters of galaxies in the neighborhoods of these more distant configurations are largely confined to the ridge or wall, in the manner of the Local Supercluster, and whether galaxies that are exceptionally luminous in the infrared are not so particularly concentrated.

\subsection{Simulations}

These observations invite comparisons with what is found in numerical simulations of cosmic structure formation. White and Silk (1979) presented an early study of the formation of the Local Supercluster modeled as a collapsing homogeneous spheroid. Neuzil, Mansfield, and  Kravtsov (2020) had far more powerful computational tools to reach their conclusion that ``Comparison with simulations shows that the number density contrasts of bright and faint galaxies within 8 Mpc alone make the Local Volume a $\simeq 2.5 \sigma$ outlier'' in the $\Lambda$CDM theory. Constrained simulations are a good start to the exploration of the formation of structures on the larger scales of the extended Local Supercluster (Yepes, Gottl{\"o}ber, and Hoffman 2014). These simulations can place mass concentrations in a near planar distribution in a satisfactory approximation to the arrangement of the local clusters of galaxies. Not yet  investigated is whether these constrained simulations also reproduce the observed scarcity of clusters at redshifts $z<0.02$ and well away from the plane. 

A useful next step in simulations might be pure dark matter simulations with resolution sufficient for identification of mass concentrations that would be expected to be clusters of galaxies if supplied with baryons, and with simulation sizes large enough to discover the rate of occurrence of arrangements of clusters similar to what is seen out to redshift $z=0.02$, and perhaps even $z\sim 0.08$. 

\subsection{Challenges}\label{sec:Challenges}

The spiral galaxy NGC~6946 looks much like the Milky Way, and it seems likely to be an equally good home for observers such as us. Since NGC~6946 is some 5~Mpc from the Local Sheet an observer in this galaxy would not so readily make out the Local Sheet of galaxies that we observe in the angular distribution of galaxies shown in Figure~\ref{fig:LU}. Are large spirals typically in sheet-like distributions, as we are, or might the sparser distribution of galaxies around NGC~6946 be more typical? 

Why are LIRGs, the galaxies that are exceptionally luminous at 12 to $100\mu$, not noticeably more common than average, or less common, near the plane of the Local Supercluster? The evidence is that these galaxies are passing through phases of rapid formation of stars and the dust that converts starlight to infrared radiation, maybe with sub-dominant AGN contributions (Tekola, V{\"a}is{\"a}nen, and Berlind 2012; Song, Linden, Evans, et al. 2021; Linden, Evans, Larson, et al. 2021; P{\'e}rez-Torres, Mattila, Alonso-Herrero et al. 2021). LIRGs tend to be in groups of galaxies, perhaps the best environment for present-day major mergers that can enhance star formation in late-type galaxies. It agrees with what is seen in the cross-correlation functions in Figures~\ref{corrlog} and~\ref{corrlinear}, which indicate that on a scale of about $0.5$~Mpc LIRGs tend to be in concentrations of 2MRS galaxies that are considerably denser than the typical concentration of 2MRS galaxies around a 2MRS galaxy. But we also see that in the region $\sim 10$~Mpc across around a typical LIRG the mean density of 2MRS galaxies is less than the mean density around a 2MRS galaxy, and only about twice the cosmic mean. We are challenged to understand why positions of LIRGs are not significantly correlated with the enhanced mass density near the plane of the Local Supercluster, and appear to be only weakly correlated with the mass distribution well away from the plane. 

A related challenge follows from the observed concentration of radio galaxies to the plane of the Local Supercluster. The evidence is that the presence of a powerful radio source in a galaxy depends on the presence of a massive central black hole. Ordinary $L\sim L_\ast$ galaxies, spirals as well as ellipticals, contain massive central black holes. So how do the black holes in radio galaxies differ from the black holes in spiral galaxies? Maybe the mass of the black hole in a radio galaxy is significantly greater than the one in an $L\sim L_\ast$ spiral. But spiral galaxies with stellar masses as large as in giant ellipticals, to judge by the luminosity at 2.2~$\mu$, might be expected to have comparable massive central black holes, yet these spirals are not noticeably concentrated to the plane of the Local Supercluster, while the similarly luminous ellipticals are concentrated to this plane, along with the radio galaxies.

Another way to look at the situation starts from the detection of the residual effects of acoustic oscillations of the plasma-radiation fluid prior to decoupling on the angular distribution of the thermal microwave background radiation (e.g. Planck Collaboration et al. 2020) and the spatial distribution of the galaxies (e. g. Beutler, Seo, Ross, et al. (2017). This is persuasive evidence that cosmic structure grew by gravity. But the situation on the smaller scales of galaxies looks more complicated. It would be natural to expect that large $L\sim L_\ast$ galaxies are more strongly clustered than $L\ll L_\ast$ dwarfs (Dekel and Silk 1986), but this is not observed. Visual examples of the similar distributions of large and small galaxies include Figure~\ref{fig:LU} in this paper, at distances $\la 6$~Mpc, and the much earlier evidence in Figures 2a and 2d in Davis, Huchra, Latham, and Tonry (1982), at distances $\la 40$~Mpc. Examples of more recent demonstrations include Einasto (1991), Binggeli (1989), and Zehavi, Zheng, Weinberg, et al. (2011). Yet we have the evidence that the most luminous early- and late-type galaxies, with similar luminosities at $2.2\mu$, and presumably similar stellar masses, have the quite different distributions relative to the plane of the Local Supercluster shown in the left-hand panels in Figure~\ref{fig:ESp}. 

The formation of the most massive galaxies seems to be bistable: early or late with occasional mixed types and irregulars. The separatrix cannot be mass; there are large and small early and late type galaxies. It cannot be angular momentum, if angular momentum is transferred by gravity and as seems likely this is not a bimodal process. We do have the hint from the evidence reviewed here that the separatrix at the high mass end is correlated with the presence of the plane of the Local Supercluster, in a manner that is difficult to attribute to gravity.

\subsection{Speculations}

These considerations invite the speculation that cosmic structure grew out of two kinds of seeds. One likely would be the usual Gaussian near scale-invariant  primordial departures from exact homogeneity, because this picture fits many tests. The other, which at $z<0.02$ would be present largely near the plane of the  growing Local Supercluster, would seed the formation of the kind of central black holes that host powerful radio sources. LIRGs would not be correlated with this plane, as observed, if they had little or nothing to do with the second kind of seed. In the $\Lambda$CDM theory clusters of galaxies form out of primeval Gaussian mass fluctuations. We do not know yet whether this Gaussian initial condition can account for the near complete concentration of clusters to the plane observed at redshifts $z\lap 0.02$. Perhaps the primeval Gaussian mass density fluctuations tend to produce groups, and the burden for formation of rich clusters of galaxies falls on the second kind of seed. 

Carr and Silk (2018) reviewed the continuing interest in the idea that primeval black holes, or the seeds for their formation, played a role in cosmic structure formation. Might they promote the formation of the kind of massive black holes that enable powerful radio sources in exceptionally luminous elliptical galaxies, while leaving the formation of ordinary galaxies with their kinds of central massive black holes to the primeval mass fluctuations that cosmological inflation is expected to produce?

Why are radio galaxies and clusters of galaxies scattered about a plane at redshift $z\lap 0.02$? One might think of the mass concentration that would be produced by a long fairly straight but kinky cosmic string drawn across space. It is difficult to imagine how the conventional picture of cosmic string formation could account for such a configuration, but in the  cyclic big bounce cosmology (Cook, Glushchenko, Ijjas, Pretorius, and Steinhardt 2020 and references therein) cosmic strings might survive the transition from contraction to expansion, stretched out by the expansion after the bounce, maybe ending up moving at a rate sufficient to pile up mass concentrations, or maybe piling up matter where the string is kinky enough to be gravitating. This is speculation, but maybe we need to think broadly. 

There are in natural science many examples of phenomena that were known well before community recognition of their significance; consider the large cosmic abundance of helium. Burbidge (1958b) recognized the evidence and the question: what produced the helium? Osterbrock and Rogerson (1961) added to the evidence, and saw that it ``can be understood without difficulty'' in Gamow's (1948) relativistic hot big bang cosmology. But community recognition came later, in 1965, prompted by the recognition that we are in a sea of microwave radiation. Are there other curiosities in the phenomenology that, when properly interpreted, will help us find a cosmology that is even better than $\Lambda$CDM? It is wishful thinking, but surely worth bearing in mind. Perhaps the extended Local Supercluster has something new and interesting to teach us. 

\section{acknowledgements}\label{sec:acknowledgements}

I am grateful for discussions with Hans B{\"o}hringer, Bernard Carr, Jaan Einasto, Merit Einasto, Ron Ekers, Yehuda Hoffman, Ken Kellerman, Andrey Kravtsov, Ofer Lahav, Peter Shaver, Paul Steinhart, Brent Tully, and Jasper Wall, and helpful comments from the referee. I acknowledge with thanks the moral support and hospitality of Princeton University, but have received no monetary support for this research. 

\section{Data Availability}

This research made use of the data compiled by the University of Hawaii  Extragalactic Distance Database; the High Energy Astrophysics Science Archive Research Center (HEASARC), which is a service of the Astrophysics Science Division at NASA/GSFC; the SIMBAD and  VizieR catalogue access tool, CDS, Strasbourg, France; the Two Micron All Sky Survey, which is a joint project of the University of Massachusetts and the Infrared Processing and Analysis Center/California Institute of Technology, funded by the National Aeronautics and Space Administration and the National Science Foundation; the NASA Astrophysics Data System Bibliographic Services; and the NASA/IPAC Extragalactic Database. I have not  generated any new data.

\label{lastpage}
\end{document}